\documentclass[aps,prb,superscriptaddress,twocolumn]{revtex4-1}
\usepackage{amsmath,amssymb,graphicx,epsfig,color,bbm}
\usepackage[pass]{geometry}
\usepackage[colorlinks=true,citecolor=blue,linkcolor=blue,urlcolor=blue,anchorcolor=black,filecolor=black]{hyperref}

\def\nn{\nonumber}

\newcommand{\ba}{\begin{eqnarray}}
\newcommand{\ea}{\end{eqnarray}}

\begin{document}

\title{Emergence of gapped bulk and metallic side walls in the zeroth Landau level in Dirac and Weyl semimetals}

\author{Ching-Kit Chan}
\affiliation{Department of Physics and Astronomy, University of California Los Angeles, Los Angeles, CA 90095, USA}
\author{Patrick A. Lee}
\affiliation{Department of Physics, Massachusetts Institute of Technology, Cambridge, MA 02139, USA}

\date{\today}

\begin{abstract}
Recent transport experiments have revealed the activation of longitudinal magnetoresistance of Weyl semimetals in the quantum limit, suggesting the breakdown of chiral anomaly in a strong magnetic field. Here we provide a general mechanism for gapping the zeroth chiral Landau levels applicable for both Dirac and Weyl semimetals. Our result shows that the zeroth Landau levels anticross when the magnetic axis is perpendicular to the Dirac/Weyl node separation and when the inverse magnetic length $l_B^{-1}$ is comparable to the node separation scale $\Delta k$. The induced bulk gap increases rapidly beyond a threshold field in Weyl semimetals, but has no threshold and is non-monotonic in Dirac systems due to the crossover between $l_B^{-1}>\Delta k$ and $l_B^{-1}<\Delta k$ regions. We also find that the Dirac and possibly Weyl systems host counterpropagating edge states between the zeroth Landau levels, leading to a state with metallic side walls and zero Hall conductance.
\end{abstract}

\maketitle

\section{Introduction}

Chiral anomaly has recently brought much excitement to condensed matter physics. One of the most striking phenomena is the negative magnetoresistance in topological semimetals in which charges are predicted to flow between two Weyl nodes in the presence of parallel electric and magnetic fields~\cite{PhysRevB.88.104412}. Since its proposal, significant experimental progress has been made to observe the chiral anomalous effect in solid state systems including Dirac~\cite{Xiong413,Li2016,Zhang2017} and Weyl~\cite{PhysRevX.5.031023,2015arXiv150302630Z} semimetals. This excitement has been pushed further by a recent transport experiment on Weyl semimetal TaAs in the extreme quantum limit~\cite{2017arXiv170406944R}. Surprisingly, it was found that the anticipated negative longitudinal magnetoresistance started to breakdown at large magnetic field ($B\sim \rm 50~T$), implying a gap opening and the loss of chiral anomaly. A second surprise is that the exponential rise in resistivity saturates at low temperature and the saturated resistivity {\it decreases} at even higher field strength ($B\sim \rm 80~T$)~\cite{2017arXiv170406944R}. Gap opening has also been suggested in other Weyl materials such as TaP~\cite{ZhangCL2017} and previous numerical studies also support similar ideas~\cite{PhysRevB.96.125153,2017arXiv170701103K}.

In this paper, we provide a generic mechanism for the field induced gap which can be commonly applied to both Dirac and Weyl semimetals. The idea is that in the presence of a magnetic field, Weyl Landau levels (LL) are formed and disperse along the magnetic axis. Because of their chiral nature, the zeroth LLs between a Weyl pair cross when the field is perpendicular to the node separation $\Delta k_W$. As discussed in Ref.~\cite{ZhangCL2017}, the crossing spectrum opens a gap when the inverse magnetic length scale $l_B^{-1}$ becomes comparable to $\Delta k_W$. We show below that this anticrossing is a consequence of the hybridization of zeroth LLs by nonlinear ladder operator couplings in the Hamiltonian. This idea can be extended to Dirac semimetals. The interesting point is that, in Dirac semimetals, Zeeman coupling provides an additional node separation scale $\Delta k_Z$. Since $\Delta k_Z \propto B$ and $l_B^{-1}\propto \sqrt B$, a transition from $\Delta k_Z < l_B^{-1}$ to $\Delta k_Z > l_B^{-1}$ takes place by increasing the field strength, giving rise to interesting field dependence of the induced gap in Dirac semimetals. Furthermore, we find that the bulk gap can support counterpropagating edge states due to the conservation of pseudospin of the Dirac Hamiltonian, so that a novel state that is metallic only on the side walls emerges.

In the following, we detail our analysis by connecting with realistic materials. Section~\ref{sect_Weyl} and section~\ref{sect_Dirac} show the zeroth LL anticrossing effects in Weyl and Dirac semimetals, respectively. Section~\ref{sect_edge} discuss the metallic side walls emerged in these two systems and then we summarize our findings in the Conclusion. Additional details about these effects and the derivations are provided in the appendix.

\section{Gapping of chiral LLs in Weyl semimetals}
\label{sect_Weyl}

To start with, consider the low-energy effective Weyl Hamiltonian
\begin{eqnarray}
H_W(\vec k)=\hbar \left( M-d k_x^2 \right) \sigma_x + \hbar v_y k_y \sigma_y + \hbar v_z k_z \sigma_z,
\label{eq_H_W}
\end{eqnarray}
which describes two Weyl points separated along $\hat k_x$ at $\vec k_{W,\chi} = (\chi \sqrt{M/d},0,0)$ with $\chi=\pm 1$ denoting the chirality. The linear energy spectrum near each Weyl point is $E_{0,\chi}(\vec k=\vec k_{W,\chi}+\vec q)  =\pm \hbar \sqrt {v_{x,\chi}^2 q_x^2+v_y^2 q_y^2+v_z^2 q_z^2} + O(q^2)$ with $v_{x,\chi}=\mp 2\sqrt{Md}$. It is understood that tilting of the Weyl cone to the type-II regime can produce an alternative gapping mechanism of the LLs~\cite{Soluyanov2015}. We focus our discussion to the type-I Weyl spectrum and do not include any tilt effect in this paper.

The chiral LLs anticross when a magnetic field is applied perpendicular to the Weyl node separation. To illustrate this, consider  $\vec B \parallel \hat k_z \perp \Delta \vec k_W$ and the gauge $\vec A = B(-y,0,0)$. Under Peierls' substitution, we have $H_W(\vec k ) \rightarrow H_{W,\perp}(\vec k) = H_W(\vec k+e\vec A/\hbar)$. $k_z$ is still a good quantum number, while $k_x$ and $k_y$ are quantized in terms of the ladder operators:
\begin{eqnarray}
k_x -\frac{eBy}{\hbar}+\xi\sqrt{\frac{M}{d}} &=& \sqrt{\frac{v_y}{|v_x|}}\frac{1}{\sqrt 2 l_B}(a^\dagger+a), \nn \\
k_y &=& \sqrt{\frac{|v_x|}{v_y}}\frac{1}{\sqrt 2 l_B}(-ia^\dagger+ia).
\end{eqnarray}
$l_B =\sqrt{\hbar/eB}$ is the magnetic length, and $a$ and $a^\dagger$ are responsible for LL quantization. $\xi$ can be $\pm 1$ and $\xi\sqrt{M/d}$ just shifts the cyclotron center. The overall Hamiltonian with $\vec B \perp \Delta
\vec k_W$ (and $\xi=+1$) becomes:
\begin{eqnarray}
H_{W,\perp}(k_z)&=&\hbar v_z k_z \sigma_z +  \frac{\hbar\sqrt{2|v_x|v_y}}{l_B}\left(\sigma_+ a + \sigma_- a^\dagger \right)  \nn \\
&&-\frac{\hbar |v_x|v_y}{8M l_B^2} \sigma_x \left( a^\dagger +a\right)^2.
\label{eq_H_perp}
\end{eqnarray}

\begin{figure}[t]
\begin{center}
\includegraphics[angle=0, width=1\columnwidth]{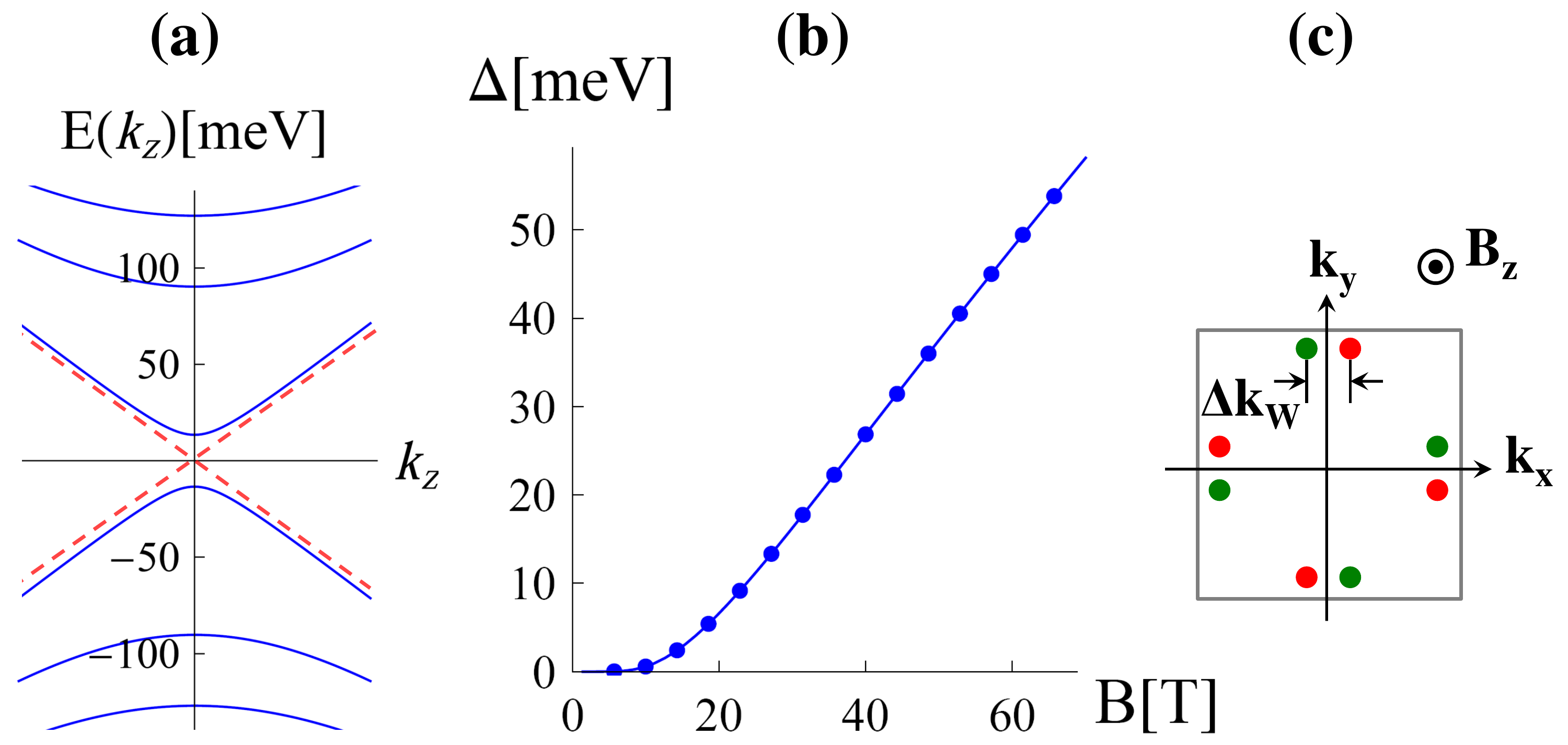}
\caption{(a) LLs of a Weyl system with two nodes separated along $\hat k_x$. $\vec B \parallel \hat k_z$ is applied perpendicular to the node separation. $\rm B=40~T$ and the system parameters are described in the text. A gap opens between the chiral LLs when $\l_B^{-1}$ is comparable to the Weyl node separation $\Delta k_W$. Dashed line depicts the crossing dispersion at small field. (b) The gap increases monotonically and non-perturbatively with the field strength. The onset field scale is $\sim O(\hbar \Delta k_W^2/e)$. (c) Schematics showing the four pairs of W1 nodes on $k_z=0$ plane in TaAs .
}
\label{fig_Weyl_gap}
\end{center}
\end{figure}

The first two terms give the LL Hamiltonian for a single Weyl point. The zeroth chiral LL state is $|\psi_{n=0} \rangle = \left|\downarrow, n=0 \right\rangle$ with the negative dispersion $E_{n=0} = -\hbar v_z k_z$. (Alternatively, we could pick $\xi=-1$ so that $\sigma_\pm \rightarrow \sigma_\mp$, after which $|\psi_{n=0} \rangle = \left|\uparrow, n=0 \right\rangle$ and $E_{n=0} = +\hbar v_z k_z$.) The last term in Eq.~(\ref{eq_H_perp}) contains nonlinear ladder operators and is responsible for the hybridization between the opposite chiral LLs. It becomes significant when ${l_B^{-1} \Delta k_{W}^{-1}}\sim O(\sqrt{|v_x|/v_y})$. For Weyl nodes with approximately isotropic Fermi velocities, this nonlinear mixing cannot be ignored when $l_B^{-1} \sim \Delta k_{W}$. Note that this nonlinear effect does not appear when the magnetic field is applied parallel to the Weyl node separation (see Appendix~\ref{sect_Weyl_parallel_field}).

Figure~\ref{fig_Weyl_gap}(a) presents the LLs of the Weyl system by diagonalizing Eq.~(\ref{eq_H_perp}) numerically.
Parameters are chosen to qualitatively resemble the band structure of TaAs~\cite{Huang2015, PhysRevX.5.011029} (see below). For small field strength, the zeroth LLs originated from the Weyl pair form a crossing spectrum with an exponentially small gap. When the field increases to $B \approx \rm 10~T$ (corresponding to $l_B^{-1}\approx 0.36 \Delta k_W$), a visible gap $\Delta$ starts to appear as shown in Fig.~\ref{fig_Weyl_gap}(b). $\Delta$ increases monotonically with $B$ within the model (with a slope $\sim e v_y/\Delta k_W$) and is non-perturbative as indicated by the activation of $\Delta$ at finite $B$ values.

We now discuss this anticrossing effect specifically for TaAs. In TaAs, there are two classes of Weyl nodes: W1 and W2. We focus on W1 nodes first and the discussion for W2 nodes will follow. There are 4 pairs of W1 nodes sitting on $k_z=0$ plane and they are related by mirror symmetry [Fig.~\ref{fig_Weyl_gap}(c)]. The two nodes in each pair are very close to each other and are separated by either $k_x=0$ or $k_y=0$ plane~\cite{Huang2015,Ma2017}. We can model two adjacent nodes with Eq.~(\ref{eq_H_W}) by taking the parameters $\hbar v_y=\rm 1.59~eV\AA$, $\hbar M=\rm 0.045~eV$ and $\hbar d=\rm 156~eV\AA^{2}$, corresponding to a node separation $\Delta k_{W}=~ 0.034~\AA^{-1}$ and an energy gap of $\rm 0.09~eV$ at $\vec k =0$. Applying $\vec B \parallel \hat k_z$ corresponds to the perpendicular field situation [Eq.~(\ref{eq_H_perp})] for all 4 Weyl pairs and consequently, gaps out all the zeroth LLs when $\l_B^{-1} \approx \Delta k_W$. Chiral anomalous effects shall break down when $\Delta/2$ is greater than the chemical potential $\mu$. Typically, $\mu \sim \rm 20-30~meV$ in TaAs, the system becomes insulating when $\Delta > 2\mu\sim \rm 40-60~meV$, corresponding to $B\sim \rm 50-70~T$ [Fig.~\ref{fig_Weyl_gap}(b)]. This threshold scale agrees qualitatively with the observed field strength above which the anomalous conductivity disappears in TaAs~\cite{2017arXiv170406944R}. A parallel analysis can be applied to the remaining 8 pairs of W2 nodes that share a similar node structure but are off the $k_z = 0$ plane. While the values for $\Delta(B)$ depends on band structure details, the anticrossing effect should be qualitatively the same.

When $\vec B \parallel \hat k_{x} $ (or $\hat k_y$), the field is perpendicular to 2 pairs of W1 nodes and parallel to the other 2 pairs. Thus, only 2 W1 pairs of zeroth LLs can be gapped. In fact, an enormous field is required to gap out the W1 nodes due to anisotropic Fermi velocities in TaAs. With $v_x \sim v_y \sim 10 v_z$ for W1 nodes~\cite{Ma2017}, a sizeable gap demands $l_B^{-1}\sim \Delta k_W \sqrt{v_x/v_z}$, meaning that the field requirement becomes ten times larger compared to the $\vec B \parallel \hat k_{z} $ case. Hence, chiral anomaly is not expected to breakdown with a realistic field strength when $\vec B \parallel \hat k_{x,y} $.

\section{LL anticrossing in Dirac semimetals}
\label{sect_Dirac}

The situation becomes very different in Dirac semimetals because of the additional momentum scale attributed by Zeeman coupling. If we ignore any Zeeman effect, the Dirac node is just a superposed copy of two Weyl nodes with opposite chirality. With two Dirac nodes separated by $\Delta k_D$, a field $\vec B \perp \Delta \vec k_D$ can gap out the zeroth LLs between the two nodes when $l_B^{-1} \approx \Delta k_D$, just like the case of Weyl semimetals. However, the presence of Zeeman coupling splits each Dirac node by a scale $\Delta k_Z \propto B $, which is much less than $\Delta k_D$ and more importantly, can be surpassed by $l_B^{-1}$ even in the weak field regime. The crossover between the two scales $\Delta k_Z$ and $l_B^{-1}$ brings in non-monotonic behaviors for the gap as shown below.

\begin{figure}[t]
\begin{center}
\includegraphics[angle=0, width=1\columnwidth]{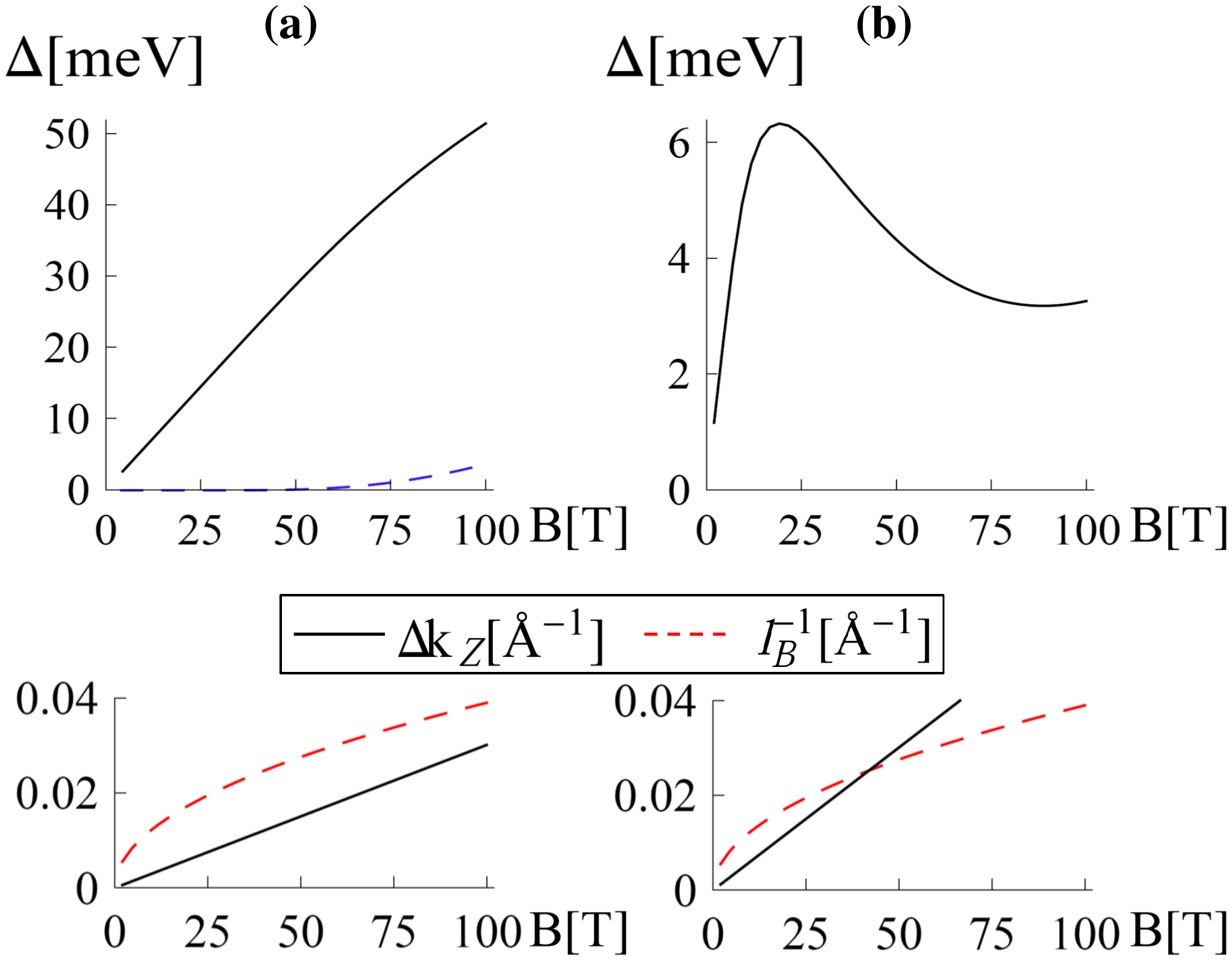}
\caption{Upper panels: (a) Gap opening between zeroth LLs in a Dirac semimetal with two Dirac nodes separated along $\hat k_z$. $\vec B \parallel \hat k_x$. Zeeman coupling splits each Dirac node by $\Delta k_Z \propto B$, which is dominated by $l_B^{-1}$ for small $B$, thus opening a gap even in the weak field limit (solid line). If Zeeman coupling is ignored, the Dirac system will behave like two copies of Weyl systems and require a large field to create a gap (dashed blue line). (b) Non-monotonic field dependence when the node separation $\Delta k_Z \propto B/\sqrt{M_0} $ is enhanced by reducing $M_0\rightarrow M_0/4$.
To explain the non-monotonic field dependence, lower panels plot the crossover between $\Delta k_Z$ and $l_B^{-1}$. The crossover points provide the field scale above which $\Delta$ no longer increases monotonically with $B$.
System parameters are detailed in the text.
}
\label{fig_Dirac_gap}
\end{center}
\end{figure}

Consider the effective four-band model applicable for Dirac semimetals such as $\rm Na_3Bi$~\cite{PhysRevB.85.195320} and $\rm Cd_3As_2$~\cite{PhysRevB.88.125427}:
\begin{eqnarray}
H_D(\vec k)
&=&
\begin{pmatrix}
M(\vec k) & v_\parallel k_+ & 0 & b^*(\vec k) \\
v_\parallel k_- & -M(\vec k) & b^*(\vec k) & 0 \\
0 & b(\vec k) & M(\vec k) & -v_\parallel k_- \\
b(\vec k) & 0 & -v_\parallel k_+ & -M(\vec k)
\end{pmatrix}.
\label{eq_H_D}
\end{eqnarray}
The Hamiltonian is expanded around the $\Gamma$ point in the basis of $| s_{\frac{1}{2}} , \frac{1}{2} \rangle $, $| p_{\frac{3}{2}} , \frac{3}{2} \rangle $, $| s_{\frac{1}{2}} , -\frac{1}{2} \rangle $ and $| p_{\frac{3}{2}} , -\frac{3}{2} \rangle $, and $M(\vec k) = M_0 - M_1 k_z^2 - M_2(k_x^2+k_y^2)$. Crystal symmetry protects each Dirac node from intermixing by enforcing $b(\vec k) \sim O(k^3)$ such that $H_D$ is block-diagonal up to $O(k^2)$. When neglecting higher order terms $b(\vec k)$, Eq.~(\ref{eq_H_D}) describes two Dirac nodes separated along $\hat k_z$ at $\vec k_{D} = (0,0,\pm \sqrt{M_0/M_1})$. Just like the Weyl semimetal analysis, we have dropped any identity term that could tilt the Dirac spectrum, since it is not important to the LL anticrossing effect.

Applying $\vec B \perp \hat k_z$, we have the Zeeman coupling:
\begin{eqnarray}
H_Z=\frac{\mu_B}{2}
\begin{pmatrix}
0 & 0 & g_s B_- & 0 \\
0 & 0 & 0 & g_p B_- \\
g_s B_+ & 0 & 0 & 0 \\
0 & g_p B_+ & 0 & 0
\end{pmatrix}.
\label{eq_H_Z}
\end{eqnarray}
$B_\pm = B_x \pm i B_y$, $\mu_B$ is the Bohr magneton and $g_{s(p)}$ is effective g-factor for the s(p) band~\cite{Xiong413,Jeon2014}. $H_Z$ splits each Dirac node along $\hat k_z$ by the scale $\Delta k_Z \sim B /\sqrt{M_0 M_1}$ (see Appendix~\ref{sect_node_shift}). Since $\vec B$ is perpendicular to the node separation, the zeroth LLs cross and open a gap depending on the ratios $l_B^{-1}/\Delta k_Z$ and $l_B^{-1}/\Delta k_D$. Below we convert the momenta to ladder operators and numerically diagonalize the total Hamiltonian.

Figure~\ref{fig_Dirac_gap}(a) plots the zeroth LL gap as a function of field strength with $\vec B \parallel \hat k_x$. We choose the parameters $M_0 =\rm -0.087~eV$, $M_1 = \rm -10.64~eV\AA^2$, $M_2 = \rm -10.36~eV\AA^2$, $v_\parallel=\rm 2.46~eV\AA$ based on band structure calculation for $\rm Na_3Bi$~\cite{PhysRevB.85.195320}, and take $g_s=18$ and $g_p=2$ according to reported values~\cite{Xiong413,Jeon2014}. Since the Dirac nodes are far apart ($\Delta k_D \sim\rm  0.18~\AA^{-1}$), it requires a large field to gap out the zeroth LLs between them [dashed line in Fig.~\ref{fig_Dirac_gap}(a)]. On the other hand, a gap immediately opens due to the strong mixing of zeroth LLs between the Zeeman split nodes. This is because the split node separation $\Delta k_Z$ is always less than $l_B^{-1}$ in the weak field limit. Different from the Weyl situation, a large field is not needed to open a sizeable $\Delta$ in Dirac semimetals. We remark that this LL anticrossing effect is different from another gapping mechanism due to the crystal rotational symmetry breaking (see Appendix~\ref{sect_rotational_symmetry}).

There is a crossover between the two scales $l_B^{-1}$ and $\Delta k_Z$ at large field. As illustrated in Fig.~\ref{fig_Dirac_gap}(b), by reducing the parameter $M_0$, which in turn enhances the node separation ($\Delta k_Z \propto B/\sqrt{M_0}$), $\Delta$ declines at large field, leading to a overall non-monotonic field dependence. This non-monotonic trend is sensitive to system parameters and it will be easier to probe in systems with a high $\Delta k_Z/B$ ratio.

\section{Edge states}
\label{sect_edge}

The creation of bulk gaps allows interesting surface states in both semimetals. Since the gaps happen between the zeroth LLs, topologically protected edge states are not expected. However, gapless edge modes are possible when the system possesses a conserved quantity. For example, spin-filtered edge states are formed in graphene in the quantum Hall (QH) regime due to the spin conservation~\cite{PhysRevLett.96.176803}. We find that, to a good approximation, such a conserved quantity exists in Dirac semimetals, leading to counterpropagating edge modes with opposite pseudospin polarizations.

\begin{figure}[t]
\begin{center}
\includegraphics[angle=0, width=1\columnwidth]{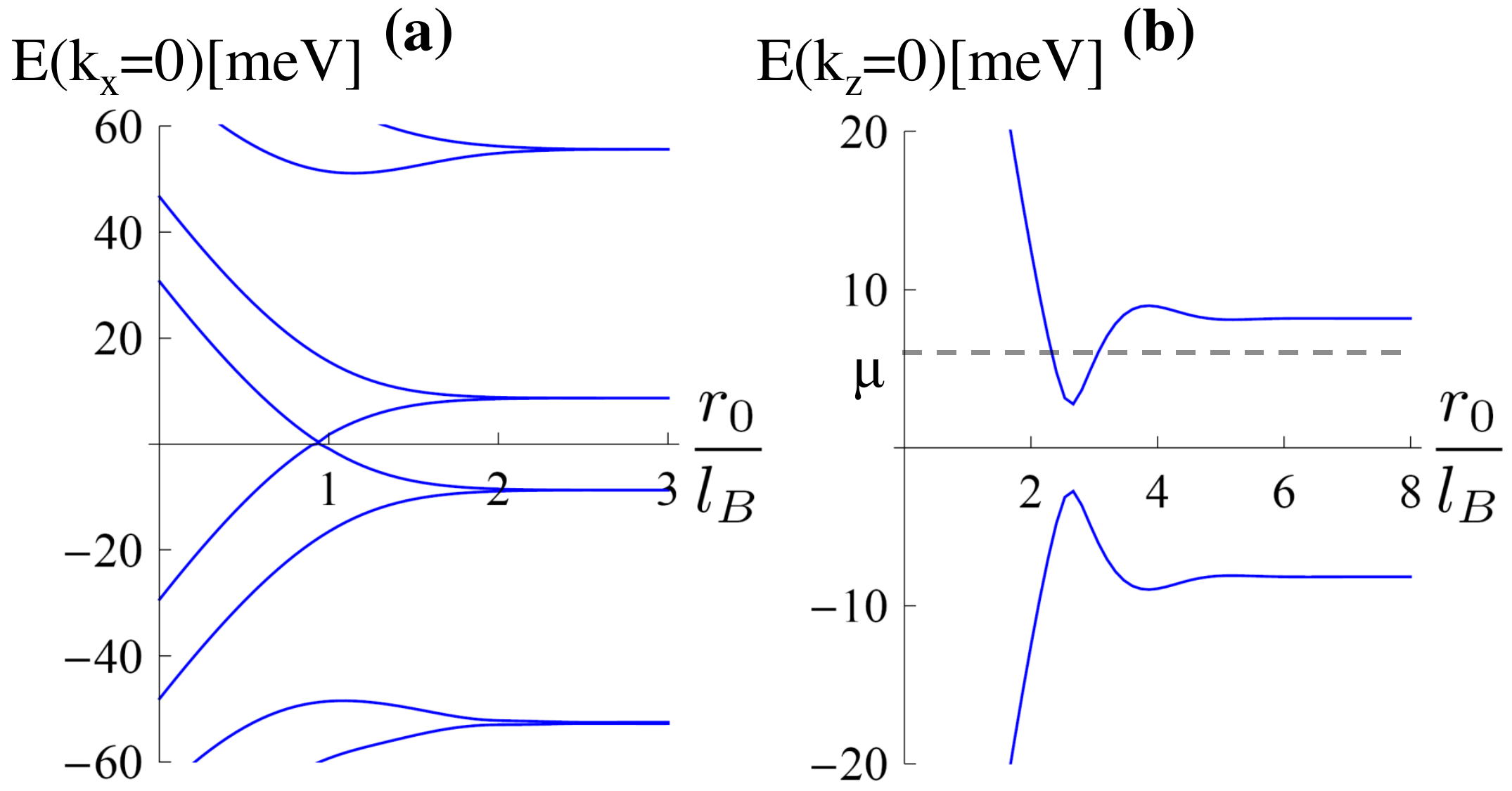}
\caption{Energy spectrum near the boundaries of the (a) Dirac and (b) Weyl systems both at $\rm B=30~T$. Same parameters used as in Fig.~\ref{fig_Dirac_gap} and Fig.~\ref{fig_Weyl_gap}. The field is perpendicular to the node separations. (a) Counterpropagating edge states between the zeroth LLs.  Up to $\rm O(k^2)$, the pseudospin of the Dirac Hamiltonian is conserved, thus allowing the formation of gapless edge modes. (b) Due to the lack of a conserved quantity in Weyl semimetals, the edge is gapped between the zeroth LLs. Depending on parameter details, the edge gap $\Delta_E$ can be smaller than the bulk gap $\Delta$ and a metallic surface is still possible in Weyl semimetals when $\Delta_E<|\mu|<\Delta$ (dashed line).}
\label{fig_edge_state}
\end{center}
\end{figure}

To explore the energy spectrum near the boundary, we augment both Hamiltonian Eq.~(\ref{eq_H_W}) and~(\ref{eq_H_D}) with a hard-wall boundary parallel to the field, just like the usual QH treatment~\cite{PhysRevB.25.2185}. In the Dirac case, we consider $\vec B\parallel \hat k_x$ and a hard-wall potential that forces the wave function to vanish for $z<0$. Using the real space representation, the Hamiltonian is diagonalized numerically as a function of $k_x$ and $k_y$. The resultant zeroth LLs form cyclotron orbits peaked around $z=r_0 = k_y l_B^2$. The Weyl system is solved in the same way by switching $k_x$ and $k_z$ axes. Note that changing the boundary orientation merely modifies the cyclotron center and does not affect our findings below.

Figure~\ref{fig_edge_state}(a) shows the Dirac LLs at $k_x=0$ near the system edge. At $\rm B=30~T$, the bulk LLs are doubly degenerate, because there are two Dirac points at different $k_z$. This degeneracy is lifted as the cyclotron center approaches the boundary, and interestingly, one upper and one lower zeroth LLs cross. This means that when $\mu<\Delta $, the surface is metallic with two counterpropagating edge modes. The edge modes persist even for a larger field strength. They are different from those lying outside the zeroth LL gap, which belong to the conventional QH edge states.

The absence of anticrossing of the counterpropagating modes originates from the conserved pseudospin of the Dirac Hamiltonian. This can be understood by rewriting Eq.~(\ref{eq_H_D}-\ref{eq_H_Z}) as $H_D+H_Z= v_\parallel k_x \Gamma_3 - v_\parallel k_y \Gamma_4 +M(\vec k) \Gamma_5 + \mu_B B_x \left[(g_s-g_p)\Gamma_{14} + (g_s+g_p)\Gamma_{23}\right]/4$, up to $\rm O(k^2)$. Since $[\Gamma_{23},H_D+H_Z]=0$ at $k_x=0$, $\Gamma_{23}=\sigma_x\tau_0$ is conserved ($\sigma$ and $\tau$ act on the pseudospin and orbital bases, respectively). The upper and lower branches of the zeroth LLs take different eigenvalues of $\Gamma_{23}$ and thus can cross near the edge. Higher order terms $b(\vec k)$ in Eq.~({\ref{eq_H_D}}) could modify the commutation relation, after which $\Gamma_{23}$ is no longer conserved and a tiny gap ($\rm < 1~meV$) could open at the edge (see Appendix~\ref{sect_node_shift}).

We can alternatively understand the protection of the gapless surface states using the mirror symmetry. The low-energy Dirac Hamiltonian has a rotational symmetry about the z-axis, which, upon the application of $B_x$, is reduced to a mirror symmetry about the yz plane. In fact, the conserved quantity $\sigma_x \tau_0$ together with $k_x \rightarrow -k_x$ constitute the mirror reflection about the yz plane. Thus, at $k_x=0$, the upper and lower zeroth LLs can be labeled by opposite mirror eigenvalues, and the corresponding surface bands can cross.

Since a Weyl system does not have a conserved quantity, the edge is generally gapped [Fig.~\ref{fig_edge_state}(b)] whose size depends on system details. However, if the Weyl points are not significantly perturbed away from the Dirac node that has a conserved quantity, the edge gap will not be as sizeable as the bulk gap. In this case, $\mu$ can reside within the bulk gap but lie outside the edge gap, leading to a metallic surface state. The evolution of the edge gap with the field can be found in the Appendix~\ref{sect_Weyl_field_evolution}.

The presence of metallic side walls could potentially explain a number of surprising observations about the magnetoresistance of TaAs~\cite{2017arXiv170406944R}. It was observed that the longitudinal resistance became thermally activated at $\rm 50~T$ but saturated at low temperature. Above $\rm 80~T$, the saturated low temperature resistivity declined. We attribute the resistivity saturation to the metallic surface state. By adjusting parameters, it is possible for our model to have an insulating bulk at an onset field $\rm 50~T$ and a surface metal between $\rm 50~T$ and $\rm 70~T$~\ref{sect_Weyl_field_evolution}. Furthermore, the surprising resistivity drop at $\rm 80~T$ could be explained by the following scenario. Suppose initially the W1 nodes have metallic surfaces and W2 nodes are gapped. Ref.~\cite{2017arXiv170406944R} shows evidence for a bulk phase transition at $\rm 80~T$, which can cause changes in parameters for W2 nodes such that the edge gap is reduced and $\mu$ moves from inside to outside the gap. This may explain the resistivity drop which is unexpected because phase transitions typically involve gap openings in the bulk.

We stress that different from topological insulators, our system is metallic only on the side walls and the top and bottom surfaces are still insulating. A conductivity measurement using Corbino geometry, i.e. attach a lead to the center of the top surface, should be able to confirm this bulk insulating property. We further remark that our surface state is not topological and is unrelated to the Fermi arc since we are in the high field regime.

The properties of our state show interesting resemblance to those in 3D QH systems by stacking QH layers~\cite{PhysRevLett.75.4496,PhysRevLett.76.2782,PhysRevLett.99.146804}. Both situations have the emergence of bulk gaps and metallic side walls. A main distinction is that our state is fully 3D and does not rely on weak interlayer couplings as in stacked QH systems. Another crucial difference is that the surface states in QH layers are chiral, which give rise to the quantized Hall conductance for each layer. Our counterpropagating surface states are not chiral and there should be no net Hall effect. In this regard, we can view our state as a 3D QH state with a Hall conductance ``quantized" to be zero.

\section{Conclusion}

Our study provides a generic and clear picture for the anticrossing effect between zeroth LLs in Dirac and Weyl semimetals. The induced gap is controlled by the ratio between $l_B^{-1}$ and Dirac/Weyl node separations. In Weyl semimetals, the gap requires a sufficiently large field, whereas in Dirac semimetals, the gap is visible even in the weak field limit. Our result provides possible explanations for the experimentally observed breakdown of chiral anomaly, the low temperature resistivity saturation and the subsequent drop of magnetoresistance in Weyl semimetals. The predicted metallic side walls with a bulk gap should be testable by conductivity measurements in Corbino geometry. Dirac semimetal is a better platform to probe these effects because the surface states are gapless due to the pseudospin conservation.

\section{Acknowledgments}

We thank Brad Ramshaw for very helpful discussions. We also thank Liang Fu for suggesting the consideration of mirror symmetry to understand the protection of gapless surface states. P.A.L. acknowledges the support from DOE Grant No. DE-FG02-03-ER46076.

\appendix
\section{Node shift and pseudospin conservation}
\label{sect_node_shift}

We provide more detail about the node shift and pseudospin conservation in Dirac semimetals. We start from a generic four band Hamiltonian respecting time reversal and inversion symmetry and then discuss the particular Hamiltonian applicable for $\rm Na_3Bi$ and $\rm Cd_3As_2$.

Consider a general $4\times 4$ matrix for a Dirac system expanded around a Dirac node with $\vec q=\vec k- \vec k_D$:
\begin{eqnarray}
H_D(\vec q)= \sum_{i=1}^{5} b_i(\vec q) \Gamma_i,
\end{eqnarray}
The representation for $\Gamma_i$ is not important regarding the node shift and conservation quantity. Here we can take $\Gamma_1 = \sigma_1 \tau_1$, $\Gamma_2 = \sigma_2 \tau_1$, $\Gamma_3 = \sigma_3 \tau_1$, $\Gamma_4 = \sigma_0 \tau_2$ and $\Gamma_5 = \sigma_0 \tau_3$, and under time reversal and inversion transformations, $\Gamma_{1,2,3,4}$ are odd and $\Gamma_5$ is even. Each $b_i(\vec q)$ can either be zero or expandable in terms of $\vec q$.

We want to find a perturbation $\Delta H$, which splits the Dirac node, and more importantly, is a conserved quantity. Specifically, suppose the node is split along $ \Delta \vec q$, we then apply an external field $\vec B \perp \Delta \vec q$, so that the zeroth Landau level is gapped in the bulk along $q_B = \vec q \cdot \vec B$. At $q_B = 0$, we require:
\begin{eqnarray}
[H_D(\vec q+e\vec A/\hbar), \Delta H]|_{q_B=0} = 0.
\label{eq_supple_commutation}
\end{eqnarray}
By doing so, at $q_B = 0$, the upper and lower zeroth Landau levels corresponds to two different eigenvalues of $\Delta H$. Since $\Delta H$ is a conserved quantity, these two bands do not mix and thus form counterpropagating gapless edge excitations.

Without loss of generosity, we can consider $\Delta H = u \Gamma_{23}$. Before we turn on the magnetic field, the eigenenergies are:
\begin{eqnarray}
E = \pm \sqrt {b_2^2 + b_3^2 + \left (\sqrt{b_1^2+b_4^2+b_5^2} \pm u \right)^2}.
\end{eqnarray}
Point or line degenerate solutions can occur at
\begin{eqnarray}
0 &=& b_2(\vec q)  \nn \\
0 &=& b_3(\vec q) \nn \\
u^2 &=&  b_1(\vec q)^2 + b_4(\vec q)^2 + b_5(\vec q)^2.
\label{eq_supple_solution}
\end{eqnarray}
When they are independent, these three equations describe a pair of Weyl node solution split from the Dirac point. If two of them are dependent and have the same solution, it corresponds to a linenode solution.

In order to satisfy Eq.~(\ref{eq_supple_commutation}) after the magnetic field is on, we have the requirement:
\begin{eqnarray}
0 &=& b_2(q_B = 0)  \nn \\
0 &=& b_3(q_B = 0),
\label{eq_supple_requirement}
\end{eqnarray}
sine other terms commute with $\Delta H$. This is possible if $b_2(\vec q)$ and $b_3(\vec q)$ are dependent and proportional to $q_B$. In this case, we have a linenode solution and $\Gamma_{23}$ is conserved. On the other hand, if $b_2(\vec q)$ and $b_3(\vec q)$ are independent, Eq.~(\ref{eq_supple_requirement}) cannot be fulfilled, the system does not have a conserved quantity and the edge will be gapped.

From the general analysis above, by perturbing a Dirac node, we either have a linenode situation with a conserved quantity and thus a gapless edge, or the Weyl splitting scenario without a conserved quantity. In the following, we explicitly consider the Dirac Hamiltonian applicable for $\rm Na_3Bi$ and $\rm Cd_3As_2$.

\begin{figure*}[t]
\begin{center}
\includegraphics[angle=0, width=1\textwidth]{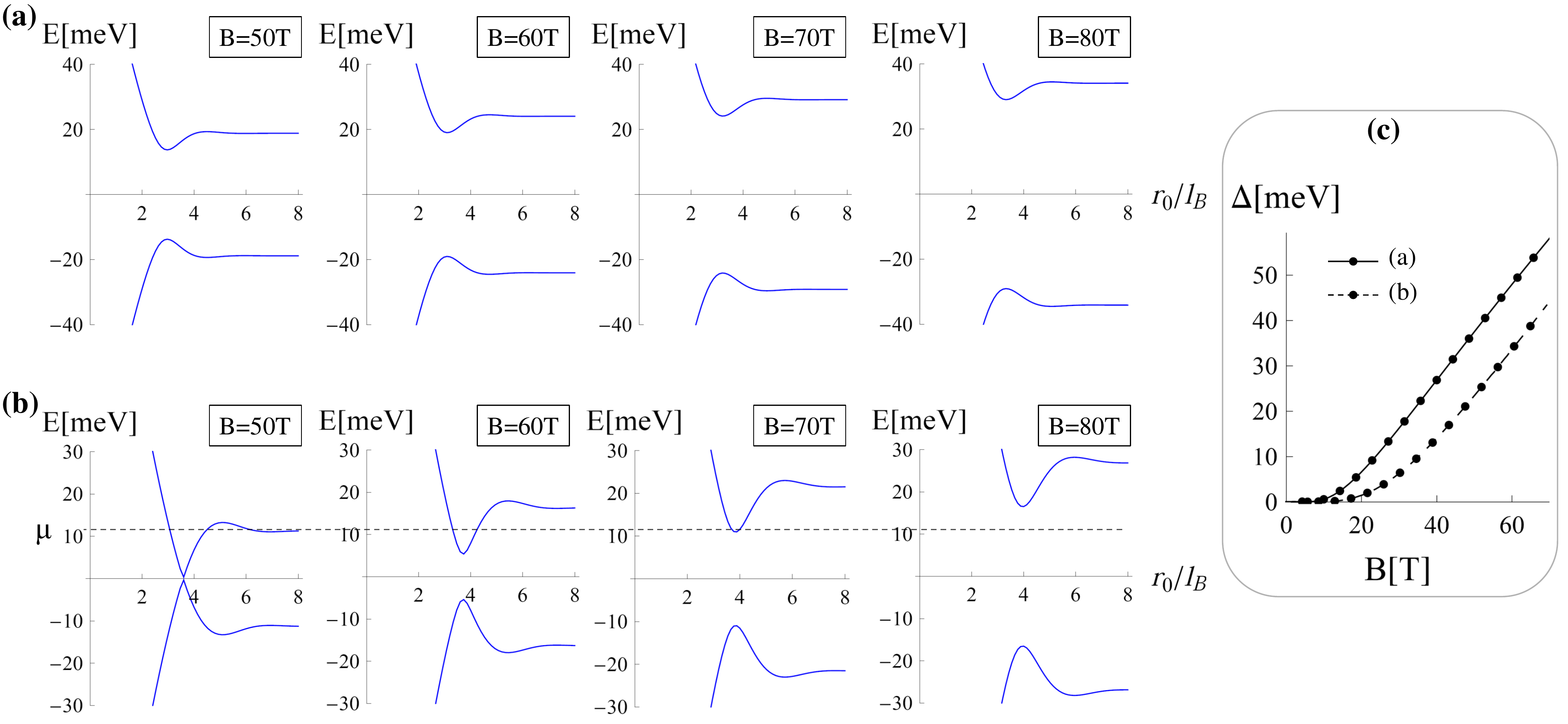}
\caption{Evolution of the edge gaps in the Weyl semimetal model using different system parameters. (a) Same parameters used in Fig.~\ref{fig_Weyl_gap} and Fig.~\ref{fig_edge_state}(b). Both edge and bulk gaps increase with the field strength above $\rm 50~T$. The metallic side walls with a bulk gap can happen but for a limited range of field strength. (b) Same Weyl node separation but doubled $v_x$. The metallic side walls can exist for a larger range of magnetic field. For example, when $\mu \rm \sim 11~meV$, the bulk becomes insulating at the onset field $\rm 50~T$ and the surface is metallic between $\rm 50~T$ and $\rm 70~T$. (c) Field dependence of the bulk gaps for these two system parameters.
}
\label{fig_Weyl_edge_field_dependence}
\end{center}
\end{figure*}

\subsection{Dirac semimetal with Zeeman perturbation}

Up to $\rm O(k^2)$, the Dirac semimetal Hamiltonian in Eq.~(\ref{eq_H_D}) has:
\begin{eqnarray}
b_i(\vec k) &=& \{0, 0, A k_x, -A k_y, M_0 - M_1 k_z^2 -M_2(k_x^2+k_y^2) \} \nn \\
&=& \{0, 0, A q_x, -A q_y, -2 \sqrt{M_0 M_1} q_z \} + O(q^2),
\end{eqnarray}
and the Zeeman perturbation due to the $B_x$ field is:
\begin{eqnarray}
\Delta H = \frac{(g_s+g_p) \mu_B B}{2} \Gamma_{23},
\end{eqnarray}
where another Zeeman term $\propto \Gamma_{14}$ is dropped for simplicity here (including it does not change the structure of the solution). Since $b_2 = 0$ is automatically satisfied, the Zeeman perturbation itself leads to a linenode solution with $q_x = 0 $ and $A^2 q_y^2 + 4 M_0 M_1 q_z^2 = (g_s+g_p)^2 \mu_B^2 B^2/4$. When we turn on the orbital coupling, we have a bulk gap between the zeroth Landau level along $q_B = q_x$. Since $\Gamma_{23}$ is conserved at $q_x=0$, the edge is gapless.

The linenode solution is an artifact due to ignoring the higher order corrections. The linenode degeneracy could be lifted and become point solutions by including $\rm O(k^3)$ corrections. For example, if we take $b_2 \propto k_y^3$, we will have point degeneracy at $k = (0,0, \pm \sqrt{(M_0 \pm (g_s+g_p)\mu_B B/2)/M_1})$, corresponding to a node shift $\Delta k_Z \propto \mu_B B/\sqrt{M_0 M_1}$. The bulk is still gapped along $q_x$. But, the edge is no longer gapless between the zeroth LLs, since $\Gamma_{23}$ is not conserved. We have checked numerically that an $\rm O(k^3)$ correction generally results in a tiny edge gap that is less than $\rm 1~meV$. The small edge gap reflects the higher order nature of the correction.

In $\rm Na_3Bi$ and $\rm Cd_3As_2$, crystal symmetry requires the $\rm O(k^3)$ corrections to take the form: $b_1 \propto k_z (k_x^2-k_y^2)$ and $b_2 \propto k_x k_y k_z$. Up to this order of correction, the conservation conditions given by Eq.~(\ref{eq_supple_requirement}) are still satisfied. $\Gamma_{23}$ is still conserved and the edge remains gapless. Higher order terms are needed to remove this conserved quantity.

\section{Alternative gap opening by breaking crystal rotational symmetry}
\label{sect_rotational_symmetry}

We contrast our LL anticrossing effect with another gapping mechanism in Dirac semimetals. The breaking of crystal rotational symmetry (such as $\rm C_3$ in $\rm Na_3Bi$~\cite{PhysRevB.85.195320} and $\rm C_4$ in $\rm Cd_3As_2$~\cite{PhysRevB.88.125427}) can couple opposite Weyl points within each Dirac node and result in massive Dirac fermions~\cite{Potter2014}. This is commonly understood in terms of the off-diagonal terms in Eq.~(\ref{eq_H_D}). By breaking the rotational symmetry, $b_{1,2}(\vec k)$ are changed from being a higher correction $\sim\beta k_z k_\pm^2$ to a linear form $\sim \beta' k_z$, which then gap out the Dirac nodes with a gap size $\sim b_{1,2}(\vec k=\vec k_D)$. In our analysis, while the magnetic field can break the rotational symmetry, the corresponding gap is rather small. To be specific, the orbital coupling leads to $b(\vec k) \rightarrow \beta k_z A_\pm^2 e^2/\hbar^2 \sim O(\beta l_B^{-3})$. Since $\beta k_D^3 \lesssim M_0 $, the gap caused by rotational symmetry breaking is estimated to be $\lesssim \frac{M_0}{k_D^3 l_B^3}$, which is about $\rm 1~meV$ at $\rm B=40~T$ and is an order of magnitude less than our $\Delta\sim \rm 20~meV$ shown in Fig.~\ref{fig_Dirac_gap}(a). In addition to the distinct gap size, our analysis is also different in that it is generally applicable to both Dirac and Weyl semimetals with and without crystal symmetry protection. Our zeroth LL gap stems from the strong hybridization of crossing chiral LLs when $l_B^{-1}$ is comparable to the Dirac/Weyl node separations.

\section{Weyl semimetal in parallel magnetic field}
\label{sect_Weyl_parallel_field}

In Section~\ref{sect_Weyl}, we show that a bulk gap is induced when the magnetic field is applied perpendicular to the Weyl node separation. On the other hand, when the magnetic axis is parallel to the node separation, the zeroth LLs do not cross and thus no gap opening is anticipated. To confirm this point, we consider the same Weyl Hamiltonian in and apply $\vec B \parallel \hat k_x$ using the gauge $\vec A = B(0,0,y)$. Under the transformations $k_y=\frac{1}{\sqrt 2 l_B}\sqrt{\frac{v_z}{v_y}}(ia^\dagger-ia)$ and $k_z+ \frac{eBy}{\hbar}=\frac{1}{\sqrt 2 l_B}\sqrt{\frac{v_y}{v_z}}(a^\dagger+a)$, we have
\begin{eqnarray}
H_{W,\parallel}(k_x)=\hbar (M-d k_x^2) \sigma_x +  \frac{\hbar\sqrt{2v_y v_z}}{l_B}\left( a \sigma_{x-}+  a^\dagger \sigma_{x+}\right), \nn \\
\label{eq_supple_H_para}
\end{eqnarray}
where $\sigma_{x\pm}=(\sigma_z\mp i \sigma_y)/2$. Contrary to the perpendicular field situation, there is no nonlinear ladder operators here. The zeroth LL is $|\psi_{n=0} \rangle = \frac{|\uparrow\rangle+|\downarrow\rangle}{2} \otimes |n=0 \rangle$ with the dispersion $E_{n=0}=    \hbar (M-d k_x^2)$ connecting the two Weyl points. Increasing the field strength merely increases the LL spacings. In this case, chiral anomaly related effect should not be affected.

\section{Field dependence of edge gaps in Weyl semimetals}
\label{sect_Weyl_field_evolution}

Here we present the field dependence of the edge gaps in Weyl semimetals. We stress that our $2\times 2$ quadratic Hamiltonian is insufficient to produce a quantitative comparison with the experimental data. Instead, we show that our model generally predicts metallic surface states for a range of magnetic field. The region of metallic surfaces is sensitive to system parameters and will certainly change if we consider a more realistic model. For the purpose of illustration, we take the simple model used in Section~\ref{sect_Weyl} and consider two different sets of parameters.

We first consider the same Weyl parameters used in the main text, i.e. $\hbar v_y=\rm 1.59~eV\AA$, $\hbar M=\rm 0.045~eV$ and $\hbar d=\rm 156~eV\AA^{2}$, giving the node separation $\Delta k_{W}=~ 0.034~\AA^{-1}$ and $v_x =\rm 5.3~eV \AA$. Figure~\ref{fig_Weyl_edge_field_dependence}(a) plots the edge gaps at different field strengths. We observe that, depending on the value of $\mu$, the edge can be metallic for a small region of field strength and will be insulating upon further increasing the magnetic field. Now, we examine a different parameter sets by increasing both $M$ and $d$ by a factor of $2$ such that the node separation remains the same but $|v_x|=2\sqrt{Md}$ is doubled. As shown in Figure~\ref{fig_Weyl_edge_field_dependence}(b), in this case, the region of surface metal is enlarged and we can have an insulating bulk and a metallic surface between $\rm 50~T$ and $\rm 70~T$. This qualitative results demonstrate the possibility of metallic surface states in Weyl semimetals for a range of magnetic field.

\bibliography{chiral-Landau-level-gap-refs}

\begin{thebibliography}{23}%
\makeatletter
\providecommand \@ifxundefined [1]{%
 \@ifx{#1\undefined}
}%
\providecommand \@ifnum [1]{%
 \ifnum #1\expandafter \@firstoftwo
 \else \expandafter \@secondoftwo
 \fi
}%
\providecommand \@ifx [1]{%
 \ifx #1\expandafter \@firstoftwo
 \else \expandafter \@secondoftwo
 \fi
}%
\providecommand \natexlab [1]{#1}%
\providecommand \enquote  [1]{``#1''}%
\providecommand \bibnamefont  [1]{#1}%
\providecommand \bibfnamefont [1]{#1}%
\providecommand \citenamefont [1]{#1}%
\providecommand \href@noop [0]{\@secondoftwo}%
\providecommand \href [0]{\begingroup \@sanitize@url \@href}%
\providecommand \@href[1]{\@@startlink{#1}\@@href}%
\providecommand \@@href[1]{\endgroup#1\@@endlink}%
\providecommand \@sanitize@url [0]{\catcode `\\12\catcode `\$12\catcode
  `\&12\catcode `\#12\catcode `\^12\catcode `\_12\catcode `\%12\relax}%
\providecommand \@@startlink[1]{}%
\providecommand \@@endlink[0]{}%
\providecommand \url  [0]{\begingroup\@sanitize@url \@url }%
\providecommand \@url [1]{\endgroup\@href {#1}{\urlprefix }}%
\providecommand \urlprefix  [0]{URL }%
\providecommand \Eprint [0]{\href }%
\providecommand \doibase [0]{http://dx.doi.org/}%
\providecommand \selectlanguage [0]{\@gobble}%
\providecommand \bibinfo  [0]{\@secondoftwo}%
\providecommand \bibfield  [0]{\@secondoftwo}%
\providecommand \translation [1]{[#1]}%
\providecommand \BibitemOpen [0]{}%
\providecommand \bibitemStop [0]{}%
\providecommand \bibitemNoStop [0]{.\EOS\space}%
\providecommand \EOS [0]{\spacefactor3000\relax}%
\providecommand \BibitemShut  [1]{\csname bibitem#1\endcsname}%
\let\auto@bib@innerbib\@empty
\bibitem [{\citenamefont {Son}\ and\ \citenamefont
  {Spivak}(2013)}]{PhysRevB.88.104412}%
  \BibitemOpen
  \bibfield  {author} {\bibinfo {author} {\bibfnamefont {D.~T.}\ \bibnamefont
  {Son}}\ and\ \bibinfo {author} {\bibfnamefont {B.~Z.}\ \bibnamefont
  {Spivak}},\ }\href {\doibase 10.1103/PhysRevB.88.104412} {\bibfield
  {journal} {\bibinfo  {journal} {Phys. Rev. B}\ }\textbf {\bibinfo {volume}
  {88}},\ \bibinfo {pages} {104412} (\bibinfo {year} {2013})}\BibitemShut
  {NoStop}%
\bibitem [{\citenamefont {Xiong}\ \emph {et~al.}(2015)\citenamefont {Xiong},
  \citenamefont {Kushwaha}, \citenamefont {Liang}, \citenamefont {Krizan},
  \citenamefont {Hirschberger}, \citenamefont {Wang}, \citenamefont {Cava},\
  and\ \citenamefont {Ong}}]{Xiong413}%
  \BibitemOpen
  \bibfield  {author} {\bibinfo {author} {\bibfnamefont {J.}~\bibnamefont
  {Xiong}}, \bibinfo {author} {\bibfnamefont {S.~K.}\ \bibnamefont {Kushwaha}},
  \bibinfo {author} {\bibfnamefont {T.}~\bibnamefont {Liang}}, \bibinfo
  {author} {\bibfnamefont {J.~W.}\ \bibnamefont {Krizan}}, \bibinfo {author}
  {\bibfnamefont {M.}~\bibnamefont {Hirschberger}}, \bibinfo {author}
  {\bibfnamefont {W.}~\bibnamefont {Wang}}, \bibinfo {author} {\bibfnamefont
  {R.~J.}\ \bibnamefont {Cava}}, \ and\ \bibinfo {author} {\bibfnamefont
  {N.~P.}\ \bibnamefont {Ong}},\ }\href {\doibase 10.1126/science.aac6089}
  {\bibfield  {journal} {\bibinfo  {journal} {Science}\ }\textbf {\bibinfo
  {volume} {350}},\ \bibinfo {pages} {413} (\bibinfo {year}
  {2015})}\BibitemShut {NoStop}%
\bibitem [{\citenamefont {Li}\ \emph {et~al.}(2016)\citenamefont {Li},
  \citenamefont {Kharzeev}, \citenamefont {Zhang}, \citenamefont {Huang},
  \citenamefont {Pletikosic}, \citenamefont {Fedorov}, \citenamefont {Zhong},
  \citenamefont {Schneeloch}, \citenamefont {Gu},\ and\ \citenamefont
  {Valla}}]{Li2016}%
  \BibitemOpen
  \bibfield  {author} {\bibinfo {author} {\bibfnamefont {Q.}~\bibnamefont
  {Li}}, \bibinfo {author} {\bibfnamefont {D.~E.}\ \bibnamefont {Kharzeev}},
  \bibinfo {author} {\bibfnamefont {C.}~\bibnamefont {Zhang}}, \bibinfo
  {author} {\bibfnamefont {Y.}~\bibnamefont {Huang}}, \bibinfo {author}
  {\bibfnamefont {I.}~\bibnamefont {Pletikosic}}, \bibinfo {author}
  {\bibfnamefont {A.~V.}\ \bibnamefont {Fedorov}}, \bibinfo {author}
  {\bibfnamefont {R.~D.}\ \bibnamefont {Zhong}}, \bibinfo {author}
  {\bibfnamefont {J.~A.}\ \bibnamefont {Schneeloch}}, \bibinfo {author}
  {\bibfnamefont {G.~D.}\ \bibnamefont {Gu}}, \ and\ \bibinfo {author}
  {\bibfnamefont {T.}~\bibnamefont {Valla}},\ }\href
  {http://dx.doi.org/10.1038/nphys3648} {\bibfield  {journal} {\bibinfo
  {journal} {Nat Phys}\ }\textbf {\bibinfo {volume} {12}},\ \bibinfo {pages}
  {550} (\bibinfo {year} {2016})}\BibitemShut {NoStop}%
\bibitem [{\citenamefont {Zhang}\ \emph
  {et~al.}(2017{\natexlab{a}})\citenamefont {Zhang}, \citenamefont {Zhang},
  \citenamefont {Wang}, \citenamefont {Liu}, \citenamefont {Chen},
  \citenamefont {Lu}, \citenamefont {Liang}, \citenamefont {Cao}, \citenamefont
  {Yuan}, \citenamefont {Tang}, \citenamefont {Li}, \citenamefont {Zhou},
  \citenamefont {Gu}, \citenamefont {Wu}, \citenamefont {Zou},\ and\
  \citenamefont {Xiu}}]{Zhang2017}%
  \BibitemOpen
  \bibfield  {author} {\bibinfo {author} {\bibfnamefont {C.}~\bibnamefont
  {Zhang}}, \bibinfo {author} {\bibfnamefont {E.}~\bibnamefont {Zhang}},
  \bibinfo {author} {\bibfnamefont {W.}~\bibnamefont {Wang}}, \bibinfo {author}
  {\bibfnamefont {Y.}~\bibnamefont {Liu}}, \bibinfo {author} {\bibfnamefont
  {Z.-G.}\ \bibnamefont {Chen}}, \bibinfo {author} {\bibfnamefont
  {S.}~\bibnamefont {Lu}}, \bibinfo {author} {\bibfnamefont {S.}~\bibnamefont
  {Liang}}, \bibinfo {author} {\bibfnamefont {J.}~\bibnamefont {Cao}}, \bibinfo
  {author} {\bibfnamefont {X.}~\bibnamefont {Yuan}}, \bibinfo {author}
  {\bibfnamefont {L.}~\bibnamefont {Tang}}, \bibinfo {author} {\bibfnamefont
  {Q.}~\bibnamefont {Li}}, \bibinfo {author} {\bibfnamefont {C.}~\bibnamefont
  {Zhou}}, \bibinfo {author} {\bibfnamefont {T.}~\bibnamefont {Gu}}, \bibinfo
  {author} {\bibfnamefont {Y.}~\bibnamefont {Wu}}, \bibinfo {author}
  {\bibfnamefont {J.}~\bibnamefont {Zou}}, \ and\ \bibinfo {author}
  {\bibfnamefont {F.}~\bibnamefont {Xiu}},\ }\href
  {http://dx.doi.org/10.1038/ncomms13741} {\bibfield  {journal} {\bibinfo
  {journal} {Nat Commun}\ }\textbf {\bibinfo {volume} {8}},\ \bibinfo {pages}
  {13741} (\bibinfo {year} {2017}{\natexlab{a}})}\BibitemShut {NoStop}%
\bibitem [{\citenamefont {Huang}\ \emph
  {et~al.}(2015{\natexlab{a}})\citenamefont {Huang}, \citenamefont {Zhao},
  \citenamefont {Long}, \citenamefont {Wang}, \citenamefont {Chen},
  \citenamefont {Yang}, \citenamefont {Liang}, \citenamefont {Xue},
  \citenamefont {Weng}, \citenamefont {Fang}, \citenamefont {Dai},\ and\
  \citenamefont {Chen}}]{PhysRevX.5.031023}%
  \BibitemOpen
  \bibfield  {author} {\bibinfo {author} {\bibfnamefont {X.}~\bibnamefont
  {Huang}}, \bibinfo {author} {\bibfnamefont {L.}~\bibnamefont {Zhao}},
  \bibinfo {author} {\bibfnamefont {Y.}~\bibnamefont {Long}}, \bibinfo {author}
  {\bibfnamefont {P.}~\bibnamefont {Wang}}, \bibinfo {author} {\bibfnamefont
  {D.}~\bibnamefont {Chen}}, \bibinfo {author} {\bibfnamefont {Z.}~\bibnamefont
  {Yang}}, \bibinfo {author} {\bibfnamefont {H.}~\bibnamefont {Liang}},
  \bibinfo {author} {\bibfnamefont {M.}~\bibnamefont {Xue}}, \bibinfo {author}
  {\bibfnamefont {H.}~\bibnamefont {Weng}}, \bibinfo {author} {\bibfnamefont
  {Z.}~\bibnamefont {Fang}}, \bibinfo {author} {\bibfnamefont {X.}~\bibnamefont
  {Dai}}, \ and\ \bibinfo {author} {\bibfnamefont {G.}~\bibnamefont {Chen}},\
  }\href {\doibase 10.1103/PhysRevX.5.031023} {\bibfield  {journal} {\bibinfo
  {journal} {Phys. Rev. X}\ }\textbf {\bibinfo {volume} {5}},\ \bibinfo {pages}
  {031023} (\bibinfo {year} {2015}{\natexlab{a}})}\BibitemShut {NoStop}%
\bibitem [{\citenamefont {{Zhang}}\ \emph {et~al.}(2015)\citenamefont
  {{Zhang}}, \citenamefont {{Xu}}, \citenamefont {{Belopolski}}, \citenamefont
  {{Yuan}}, \citenamefont {{Lin}}, \citenamefont {{Tong}}, \citenamefont
  {{Alidoust}}, \citenamefont {{Lee}}, \citenamefont {{Huang}}, \citenamefont
  {{Lin}}, \citenamefont {{Neupane}}, \citenamefont {{Sanchez}}, \citenamefont
  {{Zheng}}, \citenamefont {{Bian}}, \citenamefont {{Wang}}, \citenamefont
  {{Zhang}}, \citenamefont {{Neupert}}, \citenamefont {{Zahid Hasan}},\ and\
  \citenamefont {{Jia}}}]{2015arXiv150302630Z}%
  \BibitemOpen
  \bibfield  {author} {\bibinfo {author} {\bibfnamefont {C.}~\bibnamefont
  {{Zhang}}}, \bibinfo {author} {\bibfnamefont {S.-Y.}\ \bibnamefont {{Xu}}},
  \bibinfo {author} {\bibfnamefont {I.}~\bibnamefont {{Belopolski}}}, \bibinfo
  {author} {\bibfnamefont {Z.}~\bibnamefont {{Yuan}}}, \bibinfo {author}
  {\bibfnamefont {Z.}~\bibnamefont {{Lin}}}, \bibinfo {author} {\bibfnamefont
  {B.}~\bibnamefont {{Tong}}}, \bibinfo {author} {\bibfnamefont
  {N.}~\bibnamefont {{Alidoust}}}, \bibinfo {author} {\bibfnamefont {C.-C.}\
  \bibnamefont {{Lee}}}, \bibinfo {author} {\bibfnamefont {S.-M.}\ \bibnamefont
  {{Huang}}}, \bibinfo {author} {\bibfnamefont {H.}~\bibnamefont {{Lin}}},
  \bibinfo {author} {\bibfnamefont {M.}~\bibnamefont {{Neupane}}}, \bibinfo
  {author} {\bibfnamefont {D.~S.}\ \bibnamefont {{Sanchez}}}, \bibinfo {author}
  {\bibfnamefont {H.}~\bibnamefont {{Zheng}}}, \bibinfo {author} {\bibfnamefont
  {G.}~\bibnamefont {{Bian}}}, \bibinfo {author} {\bibfnamefont
  {J.}~\bibnamefont {{Wang}}}, \bibinfo {author} {\bibfnamefont
  {C.}~\bibnamefont {{Zhang}}}, \bibinfo {author} {\bibfnamefont
  {T.}~\bibnamefont {{Neupert}}}, \bibinfo {author} {\bibfnamefont
  {M.}~\bibnamefont {{Zahid Hasan}}}, \ and\ \bibinfo {author} {\bibfnamefont
  {S.}~\bibnamefont {{Jia}}},\ }\href@noop {} {\bibfield  {journal} {\bibinfo
  {journal} {ArXiv e-prints}\ } (\bibinfo {year} {2015})},\ \Eprint
  {http://arxiv.org/abs/1503.02630} {arXiv:1503.02630 [cond-mat.mes-hall]}
  \BibitemShut {NoStop}%
\bibitem [{\citenamefont {{Ramshaw}}\ \emph {et~al.}(2017)\citenamefont
  {{Ramshaw}}, \citenamefont {{Modic}}, \citenamefont {{Shekhter}},
  \citenamefont {{Moll}}, \citenamefont {{Chan}}, \citenamefont {{Betts}},
  \citenamefont {{Balakirev}}, \citenamefont {{Migliori}}, \citenamefont
  {{Ghimire}}, \citenamefont {{Bauer}}, \citenamefont {{Ronning}},\ and\
  \citenamefont {{McDonald}}}]{2017arXiv170406944R}%
  \BibitemOpen
  \bibfield  {author} {\bibinfo {author} {\bibfnamefont {B.~J.}\ \bibnamefont
  {{Ramshaw}}}, \bibinfo {author} {\bibfnamefont {K.~A.}\ \bibnamefont
  {{Modic}}}, \bibinfo {author} {\bibfnamefont {A.}~\bibnamefont {{Shekhter}}},
  \bibinfo {author} {\bibfnamefont {P.~J.~W.}\ \bibnamefont {{Moll}}}, \bibinfo
  {author} {\bibfnamefont {M.~K.}\ \bibnamefont {{Chan}}}, \bibinfo {author}
  {\bibfnamefont {J.~B.}\ \bibnamefont {{Betts}}}, \bibinfo {author}
  {\bibfnamefont {F.}~\bibnamefont {{Balakirev}}}, \bibinfo {author}
  {\bibfnamefont {A.}~\bibnamefont {{Migliori}}}, \bibinfo {author}
  {\bibfnamefont {N.~J.}\ \bibnamefont {{Ghimire}}}, \bibinfo {author}
  {\bibfnamefont {E.~D.}\ \bibnamefont {{Bauer}}}, \bibinfo {author}
  {\bibfnamefont {F.}~\bibnamefont {{Ronning}}}, \ and\ \bibinfo {author}
  {\bibfnamefont {R.~D.}\ \bibnamefont {{McDonald}}},\ }\href@noop {}
  {\bibfield  {journal} {\bibinfo  {journal} {ArXiv e-prints}\ } (\bibinfo
  {year} {2017})},\ \Eprint {http://arxiv.org/abs/1704.06944} {arXiv:1704.06944
  [cond-mat.str-el]} \BibitemShut {NoStop}%
\bibitem [{\citenamefont {Zhang}\ \emph
  {et~al.}(2017{\natexlab{b}})\citenamefont {Zhang}, \citenamefont {Xu},
  \citenamefont {Wang}, \citenamefont {Lin}, \citenamefont {Du}, \citenamefont
  {Guo}, \citenamefont {Lee}, \citenamefont {Lu}, \citenamefont {Feng},
  \citenamefont {Huang}, \citenamefont {Chang}, \citenamefont {Hsu},
  \citenamefont {Liu}, \citenamefont {Lin}, \citenamefont {Li}, \citenamefont
  {Zhang}, \citenamefont {Zhang}, \citenamefont {Xie}, \citenamefont {Neupert},
  \citenamefont {Hasan}, \citenamefont {Lu}, \citenamefont {Wang},\ and\
  \citenamefont {Jia}}]{ZhangCL2017}%
  \BibitemOpen
  \bibfield  {author} {\bibinfo {author} {\bibfnamefont {C.-L.}\ \bibnamefont
  {Zhang}}, \bibinfo {author} {\bibfnamefont {S.-Y.}\ \bibnamefont {Xu}},
  \bibinfo {author} {\bibfnamefont {C.~M.}\ \bibnamefont {Wang}}, \bibinfo
  {author} {\bibfnamefont {Z.}~\bibnamefont {Lin}}, \bibinfo {author}
  {\bibfnamefont {Z.~Z.}\ \bibnamefont {Du}}, \bibinfo {author} {\bibfnamefont
  {C.}~\bibnamefont {Guo}}, \bibinfo {author} {\bibfnamefont {C.-C.}\
  \bibnamefont {Lee}}, \bibinfo {author} {\bibfnamefont {H.}~\bibnamefont
  {Lu}}, \bibinfo {author} {\bibfnamefont {Y.}~\bibnamefont {Feng}}, \bibinfo
  {author} {\bibfnamefont {S.-M.}\ \bibnamefont {Huang}}, \bibinfo {author}
  {\bibfnamefont {G.}~\bibnamefont {Chang}}, \bibinfo {author} {\bibfnamefont
  {C.-H.}\ \bibnamefont {Hsu}}, \bibinfo {author} {\bibfnamefont
  {H.}~\bibnamefont {Liu}}, \bibinfo {author} {\bibfnamefont {H.}~\bibnamefont
  {Lin}}, \bibinfo {author} {\bibfnamefont {L.}~\bibnamefont {Li}}, \bibinfo
  {author} {\bibfnamefont {C.}~\bibnamefont {Zhang}}, \bibinfo {author}
  {\bibfnamefont {J.}~\bibnamefont {Zhang}}, \bibinfo {author} {\bibfnamefont
  {X.-C.}\ \bibnamefont {Xie}}, \bibinfo {author} {\bibfnamefont
  {T.}~\bibnamefont {Neupert}}, \bibinfo {author} {\bibfnamefont {M.~Z.}\
  \bibnamefont {Hasan}}, \bibinfo {author} {\bibfnamefont {H.-Z.}\ \bibnamefont
  {Lu}}, \bibinfo {author} {\bibfnamefont {J.}~\bibnamefont {Wang}}, \ and\
  \bibinfo {author} {\bibfnamefont {S.}~\bibnamefont {Jia}},\ }\href
  {http://dx.doi.org/10.1038/nphys4183} {\bibfield  {journal} {\bibinfo
  {journal} {Nat Phys}\ }\textbf {\bibinfo {volume} {advance online
  publication}} (\bibinfo {year} {2017}{\natexlab{b}})}\BibitemShut {NoStop}%
\bibitem [{\citenamefont {Borchmann}\ and\ \citenamefont
  {Pereg-Barnea}(2017)}]{PhysRevB.96.125153}%
  \BibitemOpen
  \bibfield  {author} {\bibinfo {author} {\bibfnamefont {J.}~\bibnamefont
  {Borchmann}}\ and\ \bibinfo {author} {\bibfnamefont {T.}~\bibnamefont
  {Pereg-Barnea}},\ }\href {\doibase 10.1103/PhysRevB.96.125153} {\bibfield
  {journal} {\bibinfo  {journal} {Phys. Rev. B}\ }\textbf {\bibinfo {volume}
  {96}},\ \bibinfo {pages} {125153} (\bibinfo {year} {2017})}\BibitemShut
  {NoStop}%
\bibitem [{\citenamefont {{Kim}}\ \emph {et~al.}(2017)\citenamefont {{Kim}},
  \citenamefont {{Ryoo}},\ and\ \citenamefont {{Park}}}]{2017arXiv170701103K}%
  \BibitemOpen
  \bibfield  {author} {\bibinfo {author} {\bibfnamefont {P.}~\bibnamefont
  {{Kim}}}, \bibinfo {author} {\bibfnamefont {J.~H.}\ \bibnamefont {{Ryoo}}}, \
  and\ \bibinfo {author} {\bibfnamefont {C.-H.}\ \bibnamefont {{Park}}},\
  }\href@noop {} {\bibfield  {journal} {\bibinfo  {journal} {ArXiv e-prints}\ }
  (\bibinfo {year} {2017})},\ \Eprint {http://arxiv.org/abs/1707.01103}
  {arXiv:1707.01103 [cond-mat.mtrl-sci]} \BibitemShut {NoStop}%
\bibitem [{\citenamefont {Soluyanov}\ \emph {et~al.}(2015)\citenamefont
  {Soluyanov}, \citenamefont {Gresch}, \citenamefont {Wang}, \citenamefont
  {Wu}, \citenamefont {Troyer}, \citenamefont {Dai},\ and\ \citenamefont
  {Bernevig}}]{Soluyanov2015}%
  \BibitemOpen
  \bibfield  {author} {\bibinfo {author} {\bibfnamefont {A.~A.}\ \bibnamefont
  {Soluyanov}}, \bibinfo {author} {\bibfnamefont {D.}~\bibnamefont {Gresch}},
  \bibinfo {author} {\bibfnamefont {Z.}~\bibnamefont {Wang}}, \bibinfo {author}
  {\bibfnamefont {Q.}~\bibnamefont {Wu}}, \bibinfo {author} {\bibfnamefont
  {M.}~\bibnamefont {Troyer}}, \bibinfo {author} {\bibfnamefont
  {X.}~\bibnamefont {Dai}}, \ and\ \bibinfo {author} {\bibfnamefont {B.~A.}\
  \bibnamefont {Bernevig}},\ }\href {http://dx.doi.org/10.1038/nature15768}
  {\bibfield  {journal} {\bibinfo  {journal} {Nature}\ }\textbf {\bibinfo
  {volume} {527}},\ \bibinfo {pages} {495} (\bibinfo {year}
  {2015})}\BibitemShut {NoStop}%
\bibitem [{\citenamefont {Huang}\ \emph
  {et~al.}(2015{\natexlab{b}})\citenamefont {Huang}, \citenamefont {Xu},
  \citenamefont {Belopolski}, \citenamefont {Lee}, \citenamefont {Chang},
  \citenamefont {Wang}, \citenamefont {Alidoust}, \citenamefont {Bian},
  \citenamefont {Neupane}, \citenamefont {Zhang}, \citenamefont {Jia},
  \citenamefont {Bansil}, \citenamefont {Lin},\ and\ \citenamefont
  {Hasan}}]{Huang2015}%
  \BibitemOpen
  \bibfield  {author} {\bibinfo {author} {\bibfnamefont {S.-M.}\ \bibnamefont
  {Huang}}, \bibinfo {author} {\bibfnamefont {S.-Y.}\ \bibnamefont {Xu}},
  \bibinfo {author} {\bibfnamefont {I.}~\bibnamefont {Belopolski}}, \bibinfo
  {author} {\bibfnamefont {C.-C.}\ \bibnamefont {Lee}}, \bibinfo {author}
  {\bibfnamefont {G.}~\bibnamefont {Chang}}, \bibinfo {author} {\bibfnamefont
  {B.}~\bibnamefont {Wang}}, \bibinfo {author} {\bibfnamefont {N.}~\bibnamefont
  {Alidoust}}, \bibinfo {author} {\bibfnamefont {G.}~\bibnamefont {Bian}},
  \bibinfo {author} {\bibfnamefont {M.}~\bibnamefont {Neupane}}, \bibinfo
  {author} {\bibfnamefont {C.}~\bibnamefont {Zhang}}, \bibinfo {author}
  {\bibfnamefont {S.}~\bibnamefont {Jia}}, \bibinfo {author} {\bibfnamefont
  {A.}~\bibnamefont {Bansil}}, \bibinfo {author} {\bibfnamefont
  {H.}~\bibnamefont {Lin}}, \ and\ \bibinfo {author} {\bibfnamefont {M.~Z.}\
  \bibnamefont {Hasan}},\ }\href {http://dx.doi.org/10.1038/ncomms8373}
  {\bibfield  {journal} {\bibinfo  {journal} {Nat Commun}\ }\textbf {\bibinfo
  {volume} {6}} (\bibinfo {year} {2015}{\natexlab{b}})}\BibitemShut {NoStop}%
\bibitem [{\citenamefont {Weng}\ \emph {et~al.}(2015)\citenamefont {Weng},
  \citenamefont {Fang}, \citenamefont {Fang}, \citenamefont {Bernevig},\ and\
  \citenamefont {Dai}}]{PhysRevX.5.011029}%
  \BibitemOpen
  \bibfield  {author} {\bibinfo {author} {\bibfnamefont {H.}~\bibnamefont
  {Weng}}, \bibinfo {author} {\bibfnamefont {C.}~\bibnamefont {Fang}}, \bibinfo
  {author} {\bibfnamefont {Z.}~\bibnamefont {Fang}}, \bibinfo {author}
  {\bibfnamefont {B.~A.}\ \bibnamefont {Bernevig}}, \ and\ \bibinfo {author}
  {\bibfnamefont {X.}~\bibnamefont {Dai}},\ }\href {\doibase
  10.1103/PhysRevX.5.011029} {\bibfield  {journal} {\bibinfo  {journal} {Phys.
  Rev. X}\ }\textbf {\bibinfo {volume} {5}},\ \bibinfo {pages} {011029}
  (\bibinfo {year} {2015})}\BibitemShut {NoStop}%
\bibitem [{\citenamefont {Ma}\ \emph {et~al.}(2017)\citenamefont {Ma},
  \citenamefont {Xu}, \citenamefont {Chan}, \citenamefont {Zhang},
  \citenamefont {Chang}, \citenamefont {Lin}, \citenamefont {Xie},
  \citenamefont {Palacios}, \citenamefont {Lin}, \citenamefont {Jia},
  \citenamefont {Lee}, \citenamefont {Jarillo-Herrero},\ and\ \citenamefont
  {Gedik}}]{Ma2017}%
  \BibitemOpen
  \bibfield  {author} {\bibinfo {author} {\bibfnamefont {Q.}~\bibnamefont
  {Ma}}, \bibinfo {author} {\bibfnamefont {S.-Y.}\ \bibnamefont {Xu}}, \bibinfo
  {author} {\bibfnamefont {C.-K.}\ \bibnamefont {Chan}}, \bibinfo {author}
  {\bibfnamefont {C.-L.}\ \bibnamefont {Zhang}}, \bibinfo {author}
  {\bibfnamefont {G.}~\bibnamefont {Chang}}, \bibinfo {author} {\bibfnamefont
  {Y.}~\bibnamefont {Lin}}, \bibinfo {author} {\bibfnamefont {W.}~\bibnamefont
  {Xie}}, \bibinfo {author} {\bibfnamefont {T.}~\bibnamefont {Palacios}},
  \bibinfo {author} {\bibfnamefont {H.}~\bibnamefont {Lin}}, \bibinfo {author}
  {\bibfnamefont {S.}~\bibnamefont {Jia}}, \bibinfo {author} {\bibfnamefont
  {P.~A.}\ \bibnamefont {Lee}}, \bibinfo {author} {\bibfnamefont
  {P.}~\bibnamefont {Jarillo-Herrero}}, \ and\ \bibinfo {author} {\bibfnamefont
  {N.}~\bibnamefont {Gedik}},\ }\href {http://dx.doi.org/10.1038/nphys4146}
  {\bibfield  {journal} {\bibinfo  {journal} {Nat Phys}\ }\textbf {\bibinfo
  {volume} {advance online publication}} (\bibinfo {year} {2017})}\BibitemShut
  {NoStop}%
\bibitem [{\citenamefont {Wang}\ \emph {et~al.}(2012)\citenamefont {Wang},
  \citenamefont {Sun}, \citenamefont {Chen}, \citenamefont {Franchini},
  \citenamefont {Xu}, \citenamefont {Weng}, \citenamefont {Dai},\ and\
  \citenamefont {Fang}}]{PhysRevB.85.195320}%
  \BibitemOpen
  \bibfield  {author} {\bibinfo {author} {\bibfnamefont {Z.}~\bibnamefont
  {Wang}}, \bibinfo {author} {\bibfnamefont {Y.}~\bibnamefont {Sun}}, \bibinfo
  {author} {\bibfnamefont {X.-Q.}\ \bibnamefont {Chen}}, \bibinfo {author}
  {\bibfnamefont {C.}~\bibnamefont {Franchini}}, \bibinfo {author}
  {\bibfnamefont {G.}~\bibnamefont {Xu}}, \bibinfo {author} {\bibfnamefont
  {H.}~\bibnamefont {Weng}}, \bibinfo {author} {\bibfnamefont {X.}~\bibnamefont
  {Dai}}, \ and\ \bibinfo {author} {\bibfnamefont {Z.}~\bibnamefont {Fang}},\
  }\href {\doibase 10.1103/PhysRevB.85.195320} {\bibfield  {journal} {\bibinfo
  {journal} {Phys. Rev. B}\ }\textbf {\bibinfo {volume} {85}},\ \bibinfo
  {pages} {195320} (\bibinfo {year} {2012})}\BibitemShut {NoStop}%
\bibitem [{\citenamefont {Wang}\ \emph {et~al.}(2013)\citenamefont {Wang},
  \citenamefont {Weng}, \citenamefont {Wu}, \citenamefont {Dai},\ and\
  \citenamefont {Fang}}]{PhysRevB.88.125427}%
  \BibitemOpen
  \bibfield  {author} {\bibinfo {author} {\bibfnamefont {Z.}~\bibnamefont
  {Wang}}, \bibinfo {author} {\bibfnamefont {H.}~\bibnamefont {Weng}}, \bibinfo
  {author} {\bibfnamefont {Q.}~\bibnamefont {Wu}}, \bibinfo {author}
  {\bibfnamefont {X.}~\bibnamefont {Dai}}, \ and\ \bibinfo {author}
  {\bibfnamefont {Z.}~\bibnamefont {Fang}},\ }\href {\doibase
  10.1103/PhysRevB.88.125427} {\bibfield  {journal} {\bibinfo  {journal} {Phys.
  Rev. B}\ }\textbf {\bibinfo {volume} {88}},\ \bibinfo {pages} {125427}
  (\bibinfo {year} {2013})}\BibitemShut {NoStop}%
\bibitem [{\citenamefont {Jeon}\ \emph {et~al.}(2014)\citenamefont {Jeon},
  \citenamefont {Zhou}, \citenamefont {Gyenis}, \citenamefont {Feldman},
  \citenamefont {Kimchi}, \citenamefont {Potter}, \citenamefont {Gibson},
  \citenamefont {Cava}, \citenamefont {Vishwanath},\ and\ \citenamefont
  {Yazdani}}]{Jeon2014}%
  \BibitemOpen
  \bibfield  {author} {\bibinfo {author} {\bibfnamefont {S.}~\bibnamefont
  {Jeon}}, \bibinfo {author} {\bibfnamefont {B.~B.}\ \bibnamefont {Zhou}},
  \bibinfo {author} {\bibfnamefont {A.}~\bibnamefont {Gyenis}}, \bibinfo
  {author} {\bibfnamefont {B.~E.}\ \bibnamefont {Feldman}}, \bibinfo {author}
  {\bibfnamefont {I.}~\bibnamefont {Kimchi}}, \bibinfo {author} {\bibfnamefont
  {A.~C.}\ \bibnamefont {Potter}}, \bibinfo {author} {\bibfnamefont {Q.~D.}\
  \bibnamefont {Gibson}}, \bibinfo {author} {\bibfnamefont {R.~J.}\
  \bibnamefont {Cava}}, \bibinfo {author} {\bibfnamefont {A.}~\bibnamefont
  {Vishwanath}}, \ and\ \bibinfo {author} {\bibfnamefont {A.}~\bibnamefont
  {Yazdani}},\ }\href {http://dx.doi.org/10.1038/nmat4023} {\bibfield
  {journal} {\bibinfo  {journal} {Nat Mater}\ }\textbf {\bibinfo {volume}
  {13}},\ \bibinfo {pages} {851} (\bibinfo {year} {2014})}\BibitemShut
  {NoStop}%
\bibitem [{\citenamefont {Abanin}\ \emph {et~al.}(2006)\citenamefont {Abanin},
  \citenamefont {Lee},\ and\ \citenamefont {Levitov}}]{PhysRevLett.96.176803}%
  \BibitemOpen
  \bibfield  {author} {\bibinfo {author} {\bibfnamefont {D.~A.}\ \bibnamefont
  {Abanin}}, \bibinfo {author} {\bibfnamefont {P.~A.}\ \bibnamefont {Lee}}, \
  and\ \bibinfo {author} {\bibfnamefont {L.~S.}\ \bibnamefont {Levitov}},\
  }\href {\doibase 10.1103/PhysRevLett.96.176803} {\bibfield  {journal}
  {\bibinfo  {journal} {Phys. Rev. Lett.}\ }\textbf {\bibinfo {volume} {96}},\
  \bibinfo {pages} {176803} (\bibinfo {year} {2006})}\BibitemShut {NoStop}%
\bibitem [{\citenamefont {Halperin}(1982)}]{PhysRevB.25.2185}%
  \BibitemOpen
  \bibfield  {author} {\bibinfo {author} {\bibfnamefont {B.~I.}\ \bibnamefont
  {Halperin}},\ }\href {\doibase 10.1103/PhysRevB.25.2185} {\bibfield
  {journal} {\bibinfo  {journal} {Phys. Rev. B}\ }\textbf {\bibinfo {volume}
  {25}},\ \bibinfo {pages} {2185} (\bibinfo {year} {1982})}\BibitemShut
  {NoStop}%
\bibitem [{\citenamefont {Chalker}\ and\ \citenamefont
  {Dohmen}(1995)}]{PhysRevLett.75.4496}%
  \BibitemOpen
  \bibfield  {author} {\bibinfo {author} {\bibfnamefont {J.~T.}\ \bibnamefont
  {Chalker}}\ and\ \bibinfo {author} {\bibfnamefont {A.}~\bibnamefont
  {Dohmen}},\ }\href {\doibase 10.1103/PhysRevLett.75.4496} {\bibfield
  {journal} {\bibinfo  {journal} {Phys. Rev. Lett.}\ }\textbf {\bibinfo
  {volume} {75}},\ \bibinfo {pages} {4496} (\bibinfo {year}
  {1995})}\BibitemShut {NoStop}%
\bibitem [{\citenamefont {Balents}\ and\ \citenamefont
  {Fisher}(1996)}]{PhysRevLett.76.2782}%
  \BibitemOpen
  \bibfield  {author} {\bibinfo {author} {\bibfnamefont {L.}~\bibnamefont
  {Balents}}\ and\ \bibinfo {author} {\bibfnamefont {M.~P.~A.}\ \bibnamefont
  {Fisher}},\ }\href {\doibase 10.1103/PhysRevLett.76.2782} {\bibfield
  {journal} {\bibinfo  {journal} {Phys. Rev. Lett.}\ }\textbf {\bibinfo
  {volume} {76}},\ \bibinfo {pages} {2782} (\bibinfo {year}
  {1996})}\BibitemShut {NoStop}%
\bibitem [{\citenamefont {Bernevig}\ \emph {et~al.}(2007)\citenamefont
  {Bernevig}, \citenamefont {Hughes}, \citenamefont {Raghu},\ and\
  \citenamefont {Arovas}}]{PhysRevLett.99.146804}%
  \BibitemOpen
  \bibfield  {author} {\bibinfo {author} {\bibfnamefont {B.~A.}\ \bibnamefont
  {Bernevig}}, \bibinfo {author} {\bibfnamefont {T.~L.}\ \bibnamefont
  {Hughes}}, \bibinfo {author} {\bibfnamefont {S.}~\bibnamefont {Raghu}}, \
  and\ \bibinfo {author} {\bibfnamefont {D.~P.}\ \bibnamefont {Arovas}},\
  }\href {\doibase 10.1103/PhysRevLett.99.146804} {\bibfield  {journal}
  {\bibinfo  {journal} {Phys. Rev. Lett.}\ }\textbf {\bibinfo {volume} {99}},\
  \bibinfo {pages} {146804} (\bibinfo {year} {2007})}\BibitemShut {NoStop}%
\bibitem [{\citenamefont {Potter}\ \emph {et~al.}(2014)\citenamefont {Potter},
  \citenamefont {Kimchi},\ and\ \citenamefont {Vishwanath}}]{Potter2014}%
  \BibitemOpen
  \bibfield  {author} {\bibinfo {author} {\bibfnamefont {A.~C.}\ \bibnamefont
  {Potter}}, \bibinfo {author} {\bibfnamefont {I.}~\bibnamefont {Kimchi}}, \
  and\ \bibinfo {author} {\bibfnamefont {A.}~\bibnamefont {Vishwanath}},\
  }\href {http://dx.doi.org/10.1038/ncomms6161} {\bibfield  {journal} {\bibinfo
   {journal} {Nat Commun}\ }\textbf {\bibinfo {volume} {5}},\ \bibinfo {pages}
  {5161} (\bibinfo {year} {2014})}\BibitemShut {NoStop}%
\end{thebibliography}%

\end{document}